\documentclass[aps,twocolumn,superscriptaddress,showpacs,showkeys]{revtex4-1}
\usepackage{amsmath}
\usepackage{amssymb}
\usepackage{times}
\usepackage{color}
\usepackage{epsfig,graphics}
\usepackage{bm}
\usepackage{mathrsfs}
\usepackage{enumerate}
\usepackage{indentfirst}
\usepackage{epstopdf}

\begin{document}
\preprint{APS/123-QED}

\title{New fluctuation theorem on Maxwell's demon}

\author{Qian Zeng}
\affiliation{State Key Laboratory of Electroanalytical Chemistry, Changchun Institute of Applied Chemistry, Changchun, Jilin 130022, China}
\author{Jin Wang*}
\affiliation{Department of Chemistry and of Physics and Astronomy, State University of New York, Stony Brook, NY 11794-3400 USA.}

\begin{abstract}
With the increasing interest for the control of the system at the nano and mesoscopic scales, studies have been focused on the limit of the energy dissipation in an open system by refining the concept of the Maxwell's demon. The well-known Sagawa-Ueda fluctuation theorem provides an explanation of the demon: the absence of a part of demon's information leads to an improper entropy production which violates the thermodynamic 2nd law. Realizing that the demon contributes not only to the system but also to the environments, we introduce the dissipative information to quantify the total contribution of the demon, rather than using an improper entropy production. We prove a set of new fluctuation theorems based on this, which can be used to uncover the truth behind the demon: The controlled system does not violate the 2nd law at any coarse-grained level for the demon's control. However, there exists an inevitable demon-induced dissipative information which always increases the entropy production. A consequence of these theorems is that, less work and more heat can be extracted and generated respectively by a demon than the limits predicted by the Ueda-Sagawa theorem. We also suggest a possible realization of the experimental estimation of these work and heat bounds, which can be measured and tested.
\end{abstract}

\maketitle

\section{Introduction}
In the history of physics, the well-known Maxwell's demon was proposed to act as a rebel against the authority of the thermodynamic 2nd law \cite{0}. It decreases the entropy in a thermally isolated system, and finally rescues the whole universe from the heat death. Despite the myth of its existence \cite{1,2}, the demon reflects the habitus of universal system especially at micro scales: the system interacting with a demon becomes open and thus behaves far away from thermal equilibrium. There is a deep connection between the nonequilibrium thermodynamics involved the Maxwell's demon and the information theory, such as in the much-studied cases of Szilard engine \cite{3} and Landauer principle \cite{4,5}. The physical nature of information may be revealed by the study on the demon. For this reason many efforts have been devoted to this direction. The related works have shown their importance in the theoretic and experimental areas of nano and mesoscopic system analysis and control\cite{6,7,8,9,10,11,12,13}.

As a central concept in modern thermodynamics, the \emph{entropy production} quantifies the energy dissipation and nonequilibriumness in a stochastic system. One of the fundamental properties of the entropy production is that it follows the \emph{Jarzynski equality} \cite{14} or the integral \emph{fluctuation theorem}, which is regarded as the generalized 2nd law from a microscopic perspective. To analyze the demon's effect, several pioneering works attempted to construct an \emph{improper} entropy production which disobeys the Jarzynski equality \cite{15,16,17,18,19,20}. This thought follows the original idea of Maxwell. One representative of such construction was given by Sagawa and Ueda \cite{21,22}, where a fluctuation theorem (Sagawa-Ueda theorem) has been developed for the improper entropy production by taking into account the information acquired by the demon. Correspondingly a generalized 2nd law arises from this fluctuation theorem: the demon cannot extract work more than the acquired information on average. This result gives plausible interpretation on the Szilard's engine and many other models respectively. However, there are still unsolved problems in the frameworks of such kind for the following reasons.

First, the improper entropy production arises because the system dynamics is measured in an inconsistent manner where a part of the demon's contribution is missing. Thus, the improper entropy production measures the energy dissipation and nonequilibriumness incorrectly. Intuitively the demon controls not only the system state but also the energy exchanges such as the work and heat between the system and the baths. Thus, the demon contributes to the entropies in both the system and the baths. With this thought, one can construct different improper entropy productions by neglecting any part of demon's contribution (from either the system or the baths, or parts of them). Correspondingly, there exists different fluctuation theorems for these entropy productions which can lead to different 2nd law inequalities for work or heat. The first question is which inequality is more appropriate? Second, the equality in a 2nd law inequality always represents the thermal equilibrium state of the system. However, a system is supposed to be in a nonequilibrium state when controlled by a demon. This indicates that if the demon works efficiently the equality in the 2nd law in previous frameworks does not always hold. It has been reported by several works \cite{23,24,25,26} in the examples of the information processing that the upper bound of the extracted work is less than the bound predicted by the Sagawa-Ueda theorem. This reveals the fact that when the system is at a controlled nonequilibrium state, there exists an additional energy dissipation which is not estimated by the previous frameworks. The second question is where this energy dissipation is originated from?

The motivation of this paper is to draw a clearer picture of the Maxwell's demon. We note the fact that the controlled system actually follows the 2nd law when the dynamics is properly measured. One can quantify the correct entropy productions at different coarse-grained levels for the demon's control. Every improper entropy production can give rise to a missing part of the demon's contribution. None of these entropy productions fulfills the task of complete characterization of the demon unless we take the total contribution into account. The puzzle of the demon obviously involves the interactions between the system and the demon during the whole dynamics. In the thermodynamics it is appropriate to describe these interactions by using the informational correlation -- the dynamical mutual information \cite{27,28,29}, defined as $i=\log\frac{p[x(t)|y(t)]}{p[x(t)]}$, where $x(t)$ and $y(t)$ represent the two simultaneous trajectories of the two interacting systems respectively, $p$ denotes the probability (density) of the trajectories. With this quantification at the trajectory level, it is natural to introduce the concept of \emph{dissipative information}\cite{27,28,29,30} to quantify the time-irreversibility of the dynamical mutual information,
\begin{equation}
\label{1}
  \sigma_I=i-\widetilde{i},
\end{equation}
where $\widetilde{i}=\log\frac{p[\widetilde{x}(t)|\widetilde{y}(t)]}{p[\widetilde{x}(t)]}$ is the dynamical mutual information along the time-reversed trajectories. We will show that $\sigma_I$ rightly quantifies the demon's total contribution. For a complete thermodynamical description, one should develop a set of fluctuation theorems which involves not only the entropy production in the system but also dissipative information, rather than the construction of improper entropy productions. The fluctuation theorems on the entropy productions reflect the nonequilibrium dynamics in the controlled system. Different from the ordinary fluctuation theorems for one single system, the fluctuation theorem on the dissipative information quantifies the nonequilibriumness of the interactions or binary relations. It is thus reasonable to believe that, when the demon works efficiently there exists an intrinsic nonequilibrium state (due to the binary relations) characterized by a positive averaged dissipative information. This is the source of the inevitable energy dissipation in many cases of the demon.

\section{Fluctuation Theorems and Inequalities}
Let us consider that a demon controls a system which is coupled with several thermal baths. The system and the demon are initially at the states $x_0$ and $y$ respectively. Then the demon performs a control to the system with a protocol $\Gamma(y)$ based on $y$. For simplicity the correspondence between $y$ and $\Gamma(y)$ is assumed to be bijective. Consequentially the system's trajectory $x(t)$ is correlated to the demon state $y$. As a reasonable assumption, the demon does not alter the control protocol while the demon state $y$ is unchanged during the dynamics. Driven by thermal baths, the stochasticity of the system allows the time-reversal trajectory $\widetilde{x}(t)\equiv x(\tau-t)$ to be under the identical protocol. Here the initial state of $\widetilde{x}(t)$ corresponds exactly to the final state of $x(t)$, denoted by $x_t$.

When $\Gamma(y)$ or $y$ is displayed explicitly in the system dynamics, an entropy production can be given by log ratio between the probabilities (densities) of $x(t)$ and $\widetilde{x}(t)$ conditioning on $y$,
\begin{equation}
        \label{2}
        \sigma_{X|Y}=\log\frac{p[x(t)|y]}{p[\widetilde{x}(t)|y]}=\Delta s_{X|Y}+\delta s_{X|Y}\tag{2},
\end{equation}
where the subscript $X|Y$ means that the thermodynamical entity of the system ($X$) is controlled by a given protocol of the demon ($Y$). Besides $\sigma_{X|Y}$ can be viewed as the total stochastic entropy change consisting of the contributions from the system and the baths at the microscopic level\cite{31}. This is because the total entropy change can be given by the second equality in Eq.(2). Here $\Delta s_{X|Y}=-\log p(x_t|y)-[-\log p(x_0|y)]$ quantifies the stochastic entropy difference of the system between the final and initial states; $\delta s_{X|Y}=\log\frac{p[x(t)|x_0,y]}{p[\widetilde{x}(t)|x_t,y]}$ represents the stochastic entropy flow from the system to the baths, which is also identified as the heat transferred from the baths to the system as, $Q_{X|Y}=-T\delta s_{X|Y}$, which has been proved in the detailed fluctuation theorem in the Langevin or Markovian dynamics \cite{32,33}. Thus $\delta s_{X|Y}$ is recognized as the (stochastic) entropy change in the baths. On the other hand, when the demon's control $\Gamma(y)$ or the demon state $y$ is unknown in the system dynamics, the entropy production can be measured properly at the coarse-grained level. That is to say one needs to average or integrate the demon's control information out of the dynamics, i.e., to obtain the marginal probability $p[x(t)]=\sum_y p(y)p[x(t)|y]$ with implicit control conditions. Then another entropy production, which is a coarse-grained version of $\sigma_{X|Y}$, can be given by,
\begin{equation}
        \label{3}
        \sigma_{X}=\log\frac{p[x(t)]}{p[\widetilde{x}(t)]}=\Delta s_{X}+\delta s_{X}\tag{3}.
\end{equation}
In the second equality in Eq.(3), $\Delta s_x =\log\frac{p(x_0)}{p(x_t)}$ and $\delta s_{X}=\log\frac{p[x(t)|x_0]}{p[\widetilde{x}(t)|x_t]}$ are recognized as the coarse-grained entropy changes in the system and in the baths respectively. Thus, $\sigma_{X}$ quantifies the total entropy change at the coarse-grained level with the lack of the demon's control information.An illustrative case for showing the differences between the entropy productions can be found in Fig.1. It is interesting that both $\sigma_{X|Y}$ and $\sigma_{X}$ follow the Jarzynski equalities,
\begin{equation}
        \label{4}
        \langle\exp(-\sigma_{X|Y})\rangle=1\text{, and } \langle\exp(-\sigma_{X})\rangle=1,\tag{4}
\end{equation}
where the average $\langle\exp(-\sigma_{X|Y})\rangle$ is taken over the ensembles of the system and demon's state. One should note that for every protocol, $\sigma_{X|Y}$ obeys the detailed Jarzynski equality under every possible control protocol, i.e., $\langle\exp(-\sigma_{X|Y})\rangle_{X|Y}=1$, where the average $\langle\cdot\rangle_{X|Y}$ is taken over the ensemble of the system while $y$ is fixed. For a complete view of the controlled nonequilibriumness of the system, it is appropriate to take the average of the detailed Jarzynski equality on both sides over the ensemble of demon's state, with the notation $\langle\cdot\rangle\equiv \langle\langle\cdot\rangle_{X|Y}\rangle_Y$. Notice that together the two Jarzynski equalities in Eq.(4) provide a new sight that the 2nd law holds for the system at both two levels of the knowledge of demon's control.

\begin{figure}[!ht]
\centering
\includegraphics[height=7.5cm,width=8cm]{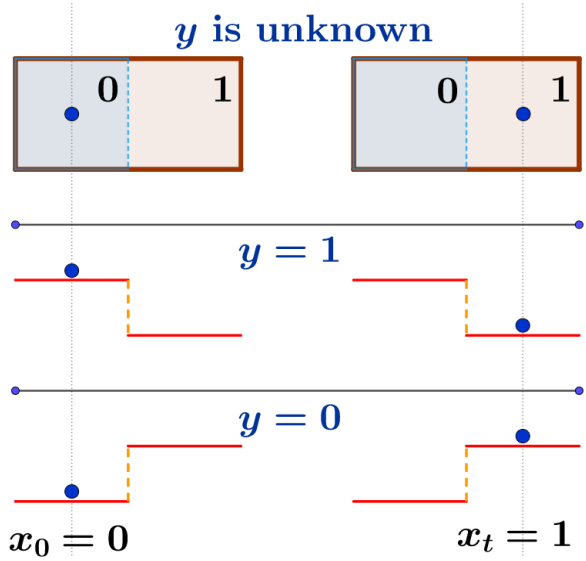}
\caption{The Entropy productions at fine and coarse-grained levels under the demon's control. A particle (shown as the blue circle) is confined in a box. The state of the particle can be represented by $0$ or $1$ when the particle is contained in the corresponding half of the box. A demon controls the particle system by exerting different potentials to the system. A trajectory of the particle in the position representation is given by $x(t)=\{x_0=0,x_t=1\}$. In the first row, the detailed information of the potential is unknown and the entropy production $\sigma_X$ can only be measured by using the coarse-grained dynamics. In the second row, the demon exerts an explicit potential to the system corresponding to $y=1$, and the  entropy production at the fine level is given by $\sigma_{X|Y=1}$ at the fine-level. In the third row, the demon exerts another potential explicitly, and the entropy production is given by $\sigma_{X|Y=0}$. The three entropy productions are not equal to each other in general.}
\label{FIG1}
\end{figure}

In general the two entropy productions shown above are different from each other. The gap between them indicates demon's contribution to entropy production, which is exactly the dissipative information $\sigma_I$ shown in Eq.(1), where the trajectory $y(t)$ is fixed at a single value of state $y$. This can be seen from the following relationship,
\begin{equation}
        \label{5.a}
        \sigma_{X|Y}=\sigma_{X}+\sigma_I\tag{5.a}.
\end{equation}
The detailed contributions of the demon to the system and the baths can be revealed by the decomposition of dissipative information, and the relations between the entropy changes shown in Eq.(2,3) in the following equalities,
\begin{equation}
        \label{5.b}
        \begin{cases}\tag{5.b}
        \sigma_I=\Delta i+\delta i\\
        \Delta s_{X|Y}=\Delta s_X+\Delta i\\
        \delta s_{X|Y}=\delta s_X+\delta i
        \end{cases}.
\end{equation}
Here $\Delta i=i_0-i_t$ is the information change of the system during the dynamics, with $i_0=\log\frac{p(x_0|y)}{p(x_0)}$ and $i_t=\log\frac{p(x_t|y)}{p(x_t)}$ being the state mutual information between the system state and demon's state at initial and final time respectively, which has been introduced in the work \cite{34,35}; $\delta i=\rho-\widetilde{\rho}$ is the time-irreversible information transfer from the demon to the system. Here $\rho=\log \frac{p[x(t)|x_0,y]}{p[x(t)|x_0]}$ and $\widetilde{\rho}=\log \frac{p[\widetilde{x}(t)|x_t,y]}{p[\widetilde{x}(t)|x_t]}$ quantify the information transferred \cite{36,37} from the demon to the system along the forward in time and backward in time trajectories respectively. It is noteworthy that the information transfer is an informational measure of how the dynamics of system depends on the demon by using the comparison between the system dynamics at different coarse-grained levels under the demon's control ($p[x(t)|x_t,y]$ and $p[x(t)|x_t]$). In Eq.(5.b), the second equality identifies the role of $\Delta i$ that it can be regarded as the demon's contribution to the entropy change in the system; the third equality indicates that $\delta i$ depicts the demon's contribution to the baths. Then the role of dissipative information is clear: it describes how the demon influences the entropy production through the nonequilibrium binary relation or interaction (see Fig.2.). Moreover, this effect can be quantified precisely in the following fluctuation theorem,
\begin{equation}
        \label{6}
        \langle\exp(-\sigma_I)\rangle=1\tag{6}.
\end{equation}
This is a new fluctuation theorem which is quite different from the Jarzynski equality because it is for the nonequilibriumness of the binary interactions between the systems rather than for a single system.

\begin{figure}[!ht]
\centering
\includegraphics[height=5cm,width=8cm]{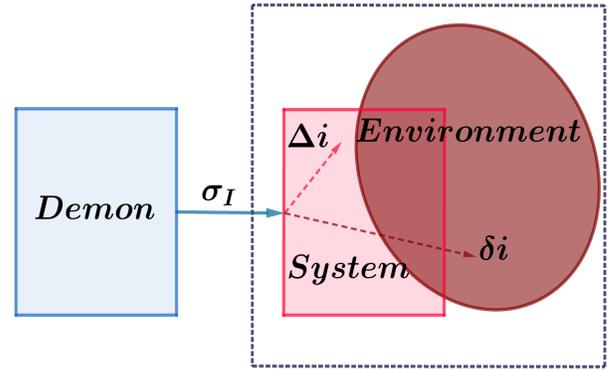}
\caption{Detailed contributions of the demon to the entropy changes in the system (denoted by $\Delta i$) and the baths (denoted by $\delta i$) respectively.}
\label{FIG2}
\end{figure}

To resolve the puzzle of the demon, we first review the construction of the improper entropy productions. A construction, $\eta=\Delta s_X+\delta s_{X|Y}$ is an improper entropy production because it violates Jarzynski equality, $\langle\exp(-\eta)\rangle\neq 1$; and in fact the two entropy changes $\Delta s_X$ and $\delta s_{X|Y}$ are measured at different levels of the knowledge of the demon's control, according to Eqs.(2,3). On the other hand, $\eta$ arises because $\Delta i$ is neglected in $\sigma_{X|Y}$, $\eta=\sigma_{X|Y}-\Delta i$, suggested by Eq.(5). This indicates that in Eq.(4) the Jarzynski equality for $\sigma_{X|Y}$ can be satisfied by adding the contribution of $\Delta i$ to $\eta$. This is the core of Sagawa-Ueda theorem which emphasizes $\Delta I=-\Delta i$ as the key characterization of the demon. Following the similar idea, one can construct different improper entropy productions. For instance, consider $\eta'=\Delta s_{X|Y} +\delta s_X$ where $\Delta s_{X|Y}$ and $\delta s_X$ are measured in an inconsistent manner in the dynamics, thus $\langle\exp(-\eta')\rangle\neq 1$. By adding $\delta i$ into $\eta'$, one has $\sigma_{X|Y}=\eta'+\delta i$, which gives rise to the same Jarzynski equality for $\sigma_{X|Y}$ in Eq.(4). However, neither $\Delta i$ nor $\delta i$ quantify the total demon's contribution, because $\Delta i$ gives the demon's influence on the system while $\delta i$ gives the demon's influence on the baths. Therefore, only the dissipative information $\sigma_I$ involving both the demon's control on the system and baths can take into account the overall contribution of the demon. Unlike previous works, the relation in Eq.(5) together with the corresponding set of fluctuation theorems in Eqs.(4,6) provides the full clear picture of Maxwell's demon.

Importantly, we further derive a series of inequalities to obtain the bounds on the dissipative entities (entropy productions and dissipative information). By applying Jensen's inequality $\langle\exp(-O)\rangle\ge \exp(-\langle O\rangle)$ to Eqs.(4,6) respectively, we have
\begin{equation}
        \label{7}
\begin{cases}\tag{7}
        \langle\sigma_{X|Y}\rangle\ge0,\text{ or } \langle\Delta s_{X|Y}\rangle\ge-\langle\delta s_{X|Y}\rangle\\
        \langle\sigma_X\rangle\ge0,\text{ or }\langle\Delta s_{X}\rangle\ge-\langle\delta s_{X}\rangle\\
        \langle\sigma_I\rangle\ge0,\text{ or } \langle\Delta i\rangle\ge-\langle\delta i\rangle
\end{cases}
\end{equation}
The first two are the 2nd law inequalities at different coarse-grained levels of the demon's control corresponding to the Jarzynski equalities in Eq.(4), while the last inequalities about $\sigma_I$ shows the new feature of the nonequilibrium behavior brought by the demon. To see this, take the average on both sides of Eq.(5) over the ensembles, we have $\langle\sigma_{X|Y}\rangle=\langle\sigma_X\rangle+\langle\sigma_I\rangle$. Combining with Eq.(7), one sees that $\langle\sigma_{X|Y}\rangle$ quantifies the true (utmost) entropy productions in the system. A lower bound of $\langle\sigma_{X|Y}\rangle$ different from that obtained from the 2nd law in Eq.(7) (which is zero) is given by the following inequality,
\begin{equation}
        \label{8}
        \langle\sigma_{X|Y}\rangle\ge\langle\sigma_I\rangle\ge0\tag{8}.
\end{equation}
$\langle\sigma_{X|Y}\rangle=0$ at the fine level indicates that the system is in a quasi-static (equilibrium) process where every control protocol is applied infinitely slowly. Such a demon does not work efficiently in practice. High efficiency means achieving the control in a finite time, which leads to a nonequilibrium process. Consequentially the lower bound of  $\langle\sigma_{X|Y}\rangle$ is always a positive number rather than $0$. Although measured properly, $\langle\sigma_X\rangle$ does not reflect the true nonequilibriumness of the system due to the coarse-graining. Meanwhile, $\langle\sigma_X\rangle$ does not need to be strictly positive when the system is actually in nonequilibrium. However, there always exists a positive dissipative information ($\langle\sigma_I\rangle>0$) which is contained in the true entropy production, $\langle\sigma_{X|Y}\rangle$. This is due to the nonequilibrium part of the dynamical mutual information for the binary relationship between the demon and the system. The exception can be seen in the case where a demon controls the system with an unique and deterministic protocol, we have $\langle\sigma_I\rangle=0$ as $\langle\sigma_X\rangle=\langle\sigma_{X|Y}\rangle$ during the dynamics. Otherwise, there exists an intrinsic nonequilibrium state of the system in general, which is characterized by an inevitable energy dissipation given by $\langle\sigma_{X|Y}\rangle=\langle\sigma_I\rangle>0$.

\section{New Bounds for Work and Heat}
A consequence of Eq.(8) is that the bounds on the heat and work should be revised beyond the ordinary 2nd law. To see this, let us assume that the system is coupled with a thermal bath with temperature $T$ for simplicity. Then the system dynamics can be given by the Langevin dynamics. The Hamiltonian of the system depends on the system state and the control protocol, denoted by $H(x,y)\equiv H(x,\Gamma(y))$ (correspondence between $y$ and $\Gamma(y)$ is bijective). The change in the Hamiltonian during the dynamics can be given by $\Delta H_{X|Y}=H(x_t,y)-H(x_0,y)$. With the assumption of the detailed FT, the entropy production can be given in terms of the heat absorbed by the system, $\sigma_{X|Y}=\Delta s_{X|Y}-T^{-1}Q_{X|Y}$. According to the thermodynamic 1st law, $\Delta H_{X|Y}=Q_{X|Y}+W_{X|Y}$ where $W_{X|Y}$ is the work performed on the system, $\sigma_{X|Y}$ can be rewritten in terms of the work, $\sigma_{X|Y}=T^{-1}(W_{X|Y}-\Delta F_Y)$. Here $\Delta F_Y$ is the Helmholtz free energy difference, $\Delta F_Y=\langle \Delta H_{X|Y}\rangle_{X|Y}-T\langle \Delta s_{X|Y}\rangle_{X|Y}$, under certain protocol. The probability weights in the averages of the state variables in $\Delta F_Y$ should be distinguished at the initial and final states: the weights $p(x_0|y)$ and $p(x_t|y)$ are used for $x_0$ and $x_t$ respectively. Then according to the inequality for $\sigma_{X|Y}$ in Eq.(7), we reach the ordinary 2nd law inequalities for the heat and work,
\begin{equation}
\label{9}
   \langle Q_{X|Y} \rangle\le T \langle\Delta s_{X|Y}\rangle \text{, and } \langle W_{X|Y}\rangle\ge \Delta F\tag{9}.
\end{equation}
Here, $\Delta F=\langle \Delta F_Y\rangle_{Y}$ is recognized as the averaged free energy difference over the ensemble of the demon state. It is noteworthy that different constructions of improper entropy productions may lead to different forms of Eq.(9). However, by noting the relation in Eq.(5) and rearranging terms, they are equivalent to each other. Because all the improper entropy productions are generated by decomposing $\sigma_{X|Y}$ in different ways. On the other hand, we take the dissipative information into account. By noting Eq.(8), we reach tighter bounds for the heat and the work, compared to Eq.(9),
\begin{equation}
\label{10}
\begin{matrix}\tag{10}
   &\langle Q_{X|Y}\rangle\le T \langle\Delta s_{X|Y}\rangle-T \langle\sigma_{I}\rangle\le T\langle\Delta s_{X|Y}\rangle;&\\
   &\langle W_{X|Y}\rangle\ge \Delta F+T \langle\sigma_{I}\rangle\ge \Delta F. &
\end{matrix}
\end{equation}
Here we obtain a smaller upper bound for the heat and a larger lower bound for the work than the ordinary 2nd law. These new tighter bounds clearly indicate the nontrivial nonequilibrium state of a system controlled by a demon. It is important to note that, compared to the tighter bounds in the first inequalities in Eq.(10), the looser bounds of the heat and work in the second inequalities (also see Eq.(9)), which are also predicted by the Sagawa-Ueda theorem, represent the equilibrium limit while the demon does not work efficiently.

Usually in the practical model of Maxwell's demon such as the Szilard's type demon, the action of the demon is divided into two different processes: measurement and feedback control. In the measurement process, the demon observes the system and acquires the information of the system state. The demon is usually implemented by a physical system, and the measured system can be viewed as the outer controller of the demon. In this situation, an inevitable heat, or say, the measurement heat $Q_{mea}$ can be generated from the demon during the information acquirement \cite{38,39}. In the feedback control process, the demon extracts a positive work $W_{ext}$ from the system with an additional energy dissipation. By noting the relations $Q_{mea}=-Q_{X|Y}$ and $W_{ext}=-W_{X|Y}$, the bounds for $Q_{mea}$ and $W_{ext}$ can be given by the ordinary 2nd law in Eq.(9) where the equalities hold for infinitely slow quasi-static or equilibrium processes. However, if the demon works efficiently, we then come to a nonequilibrium situation where an positive energy dissipation is originated from the dissipative information. Thus, new bounds for $Q_{mea}$ and $W_{ext}$ can be given by Eq.(10) such that
\begin{equation}
\label{11}
\begin{matrix}\tag{11}
   &\langle Q_{mea}\rangle\ge T \langle\sigma_{I}\rangle-T \langle\Delta s_{X|Y}\rangle\ge -T\langle\Delta s_{X|Y}\rangle;&\\
   &\langle W_{ext}\rangle\le -\Delta F-T \langle\sigma_{I}\rangle\le -\Delta F &
\end{matrix}
\end{equation}
This means that there is more heat generated in the measurement and less work extracted in the feedback control than the estimations given by the ordinary 2nd law.

\section{Illustrative Cases}
To illustrate our idea in this letter, we calculate the cases of the information ratchets shown in Fig.3, which can be tested in the experiments. A potential with the two wells is exerted on a confined particle. The height between the two wells is equal to $V>0$. While under the equilibrium, the probabilities that the particle is at the lower and the higher well can be quantified by $p_l=[1+\exp(-V)]^{-1}$ and $p_h=1-p_l$ respectively ($p_l>1/2$). An outside controller can control the particle by reversing the profile of the potential, i.e., by raising the lower well up to $V$ and lowering the higher well down to $0$. The action of the controller is assumed to be fast enough before the particle reacts. For no loss of generality, the temperature of the environmental bath is assumed to be at $T=1$.

\begin{figure}[!ht]
\centering
\includegraphics[height=2.5cm,width=8cm]{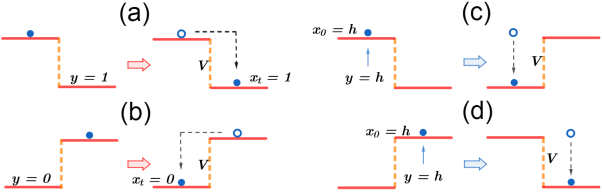}
\caption{The confined particle works as a demon or is controlled by a demon. In (a) and (b), the particle is used as a demon and measures the system state before the control. The system state, denoted by $y$, is represented by the location of the lower well ($0$ or $1$) in the potential. In (a), $y=1$; and In (b), $y=0$. The final state of the particle, denoted by $x_t$, is taken as the state of $y$. In (c) and (d), the particle is controlled by a demon. The initial state of the particle, denoted by $x_0$, is represented by $h$ or $l$ when the particle is at the higher or the lower well. When spotting the state $x_0=h$, the demon reverses the potential and extracts a positive work of $V$ from the particle system.}
\label{FIG3}
\end{figure}

If the particle works as a demon (shown in Fig.3. (a) and (b)), the particle is supposed to measure the state of the controlled system at first. The state of the particle is denoted by $x=0$ or $1$ when being at the left or the right well respectively. The system state can be represented by the location of the lower well with the value of $y=0$ or $1$ with equal probability $p(y=0)=p(y=1)=1/2$. The particle is initially under the equilibrium until the system state changes. Correspondingly, the potential is reversed by the system immediately and the particle starts measuring the current system state. When the equilibrium is achieved, the final state $x_t$ of the particle is taken as an observation of $y$. The probability of the measurement error can be given by the probability of the particle at the higher well, $p_{X_t|Y}(x_t\neq y|y)=p_h$. On the other hand, the measurement precision is characterized by the probability of the particle at the lower well, $p_{X_t|Y}(x_t= y|y)=p_l$. By noting the definitions and relationships shown in Eq.(2), the averaged measurement heat generated by the particle can be given by $\langle Q_{mea}\rangle=(1/2-p_h)V$, and the entropy change can be evaluated by $\langle\Delta s_{X|Y}\rangle=-I_t$ (see Eqs.(S.26-S.28) in \cite{40}). Here $I_t=\log 2-S\ge0$ is the final mutual information which measures the correlation between the observation $x_t$ and the state $y$ \cite{34,35}, where the Shannon entropy $S$ is given by $S=-p_l\log p_l-p_h\log p_h$. Then according to the Eq.(11), the new bound of $\langle Q_{mea}\rangle$ in this case can be given by
\begin{equation}
\label{12}
\langle Q_{mea}\rangle\ge \langle\sigma_{I}\rangle+I_t \ge I_t.\tag{12}
\end{equation}
Here the dissipative information $\langle\sigma_{I}\rangle$ can be calculated by using the probabilities of the forward and backward trajectories $x(t)=\{x_0,x_t\}$ and $\widetilde{x}(t)=\{x_t,x_0\}$ respectively. By inserting these probabilities into Eq.(1), we have the expression $\langle\sigma_{I}\rangle=\log\sqrt{2p_l^2+2p_h^2}\ge0$ (see Eq.(S.29) in \cite{40}). Although the measurement precision characterized by $p_l$ increases as the potential height $V$ increases, higher precision also raises up both the averaged measurement heat and the lower bound of the energy dissipation quantified by the dissipative information in this case. The numerical results can be found in Fig.4. (a) and (b).

\begin{figure}[!ht]
\centering
\includegraphics[height=5cm,width=8cm]{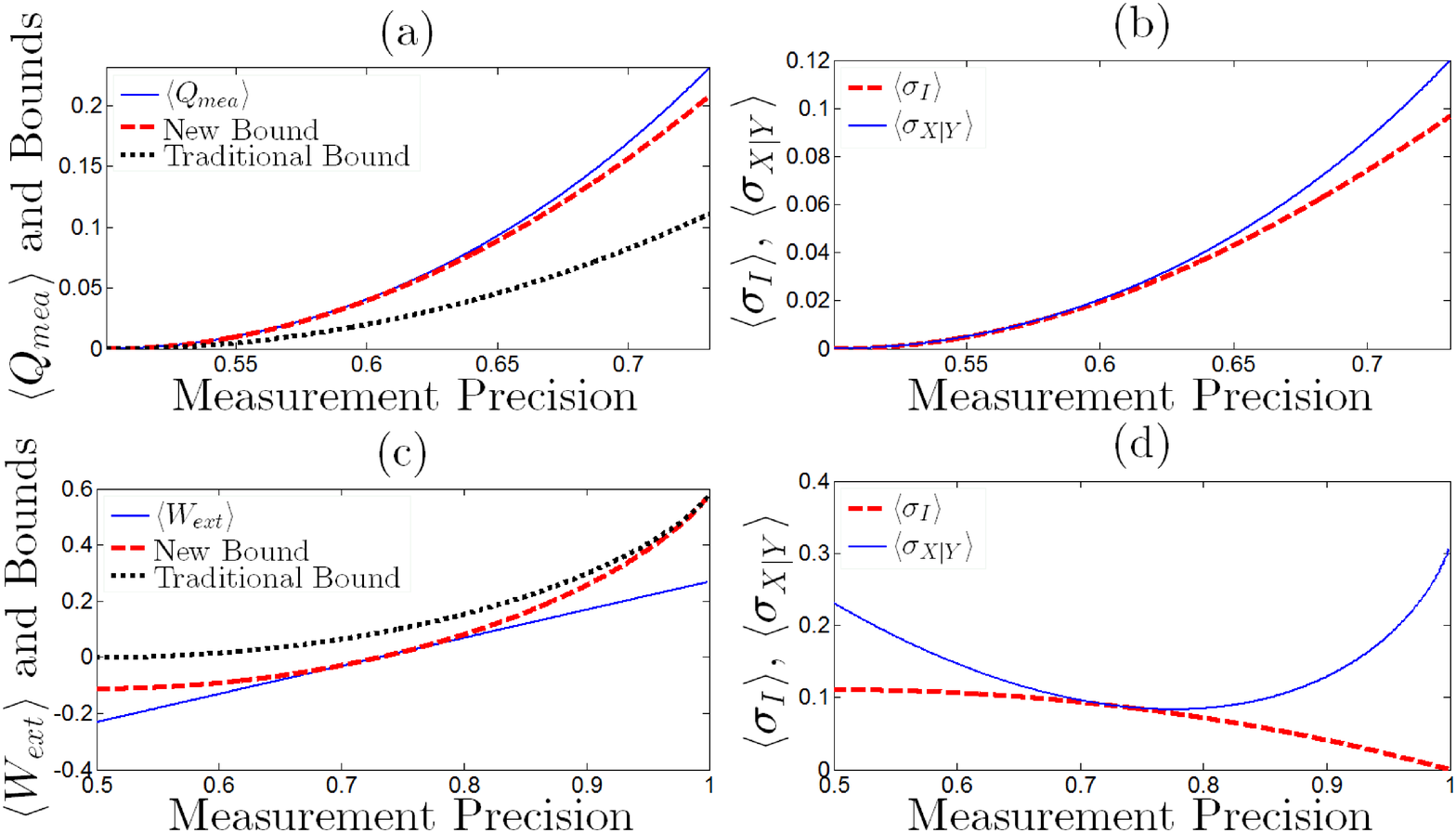}
\caption{In (a), the averaged measurement heat $\langle Q_{mea}\rangle$ (solid line), the traditional lower bound $I_t$ (dotted line), and the new lower bound $I_t+\langle\sigma_I\rangle$ (dash line) in the measurement are plotted as functions of the measurement precision $p_l$. The potential height $V$ is raised from $0$ to $1$. Correspondingly, $p_l$ is increased from $0.50$ to $0.73$ monotonically. The corresponding dissipative information $\langle\sigma_I\rangle$ (dash line) and the entropy production $\langle\sigma_{X|Y}\rangle$ (solid line) in the measurement are shown as the functions of the precision $p_l$ in (b). In (c), the extracted work $\langle W_{mea}\rangle$ (solid line), the traditional upper bound $I_0$ (dotted line), and the new upper bound $I_0-I_c$ (dash line) are plotted as functions of the measurement precision $1-\epsilon$, where $1-\epsilon$ is ranged from $0.5$ to $1$ and the potential height is $V=1$. The corresponding dissipative information $\langle\sigma_I\rangle$ (dash line) and the entropy production $\langle\sigma_{X|Y}\rangle$ (solid line) are shown in (d).}
\label{FIG4}
\end{figure}

In the next, we use a demon to extract positive work from the particle system (shown in Fig.3. (c) and (d)). In this case, the state of the particle can be denoted by $x=l$ or $h$ when at the lower or the higher well respectively. Initially, the particle is under equilibrium. The demon measures the state of the particle at first and obtains the observation $y$. The demon plays the feedback control according to the observation $y$. When the particle is observed to be at the higher well, the demon reverses the potential immediately and extracts an amount of work $W_{ext}=V$. After the control, the demon does nothing until the particle goes to the equilibrium again. For a practical thought, the demon's measurement can have a random error and this error certainly lowers the efficiency of the work extraction. Here we simply assume that the measurement error occurs with stable probability $p_{Y|X_0}(y\neq x_0|x_0)=\epsilon$. By using Eq.(2) and noting the thermodynamical 1st law, the extracted work can be given by $\langle W_{ext}\rangle=(p_h-\epsilon)V$ on average, and the efficient free energy difference is equal to the mutual information change during the dynamics, $\Delta F=-I_0$ (see Eqs.(S.30-S.33) in \cite{40}). Here the mutual information $I_0=S_Y-S_\epsilon\ge0$ represents the initial correlation between the demon and the particle, where the Shannon entropies can be given by $S_Y=-p_y\log p_y-(1-p_y)\log(1- p_y)$ and $S_\epsilon=-\epsilon\log \epsilon-(1-\epsilon)\log (1-\epsilon)$, with $p_y=p_l(1-\epsilon)+p_h\epsilon$ representing the probability of the observation $y=l$. Then due to Eq.(11), the new bound of $\langle W_{ext}\rangle$ can be given by
\begin{equation}
\label{13}
\langle W_{ext}\rangle\le I_0-I_c \le I_0,\tag{13}
\end{equation}
Here the mutual information $I_c=S_Y-S\ge0$ measures correlation between the demon and the particle right after the control, where the Shannon entropy is $S=-p_l\log p_l-p_h\log p_h$. It is important to note that $I_c$ is actually the information that is not used to extract the work but can merely dissipate into the bath as the dissipative information $\langle\sigma_{I}\rangle=I_c$. This result can be verified by evaluating Eq.(1) (see Eq.(S.18) in \cite{40}). For this reason, the demon can only extract work less than the mutual information difference before and after the control, quantified by $I_0-I_c$. We can see that higher measurement precision characterized by $1-\epsilon$ can increase the averaged extracted work (with fixed potential height $V$), in the meanwhile the inevitable dissipative information is decreased by the increasing precision in this case. The numerical results are shown in Fig.2. (c) and (d). Also, we can note that the dissipative information $\langle \sigma_{I}\rangle$ bounds the entropy production $\langle \sigma_{X|Y}\rangle$ from the below in both the cases of the measurement and feedback control (see Fig.4. (b) and (d)). This verifies the inequality in Eq.(8).

On the other hand, we find that the tighter upper bound in Eq.(13) is equivalent to the \emph{information process 2nd law}\cite{24,25,26}, by noting $I_0-I_c=S-S_\epsilon$. Here $S$ and $S_\epsilon$ can be regarded as the Shannon entropies of a ``0,1" tape before and after the information processing respectively\cite{23}. This indicates that the proposed FTs in this letter can be applied to the area of thermodynamics computing from a general perspective. In addition, the looser bounds for the heat and work in Eqs.(12,13) are predicted by the 2nd law (Sagawa-Ueda theorem), and these bounds can only be achieved in the quasistatic (or equilibrium) control protocols.

\section{Conclusion}
Traditional analysis on the Maxwell's demon focus on how the 2nd law is violated by the system, and is rescued by some hidden demon-induced entities. These entities were believed as the key characterization of the demon. In contrast, we show that the system does not disobey the 2nd law whether the demon is hidden or not, which can be seen in the new set of fluctuation theorems (Eq.(4)) for the entropy productions when they are correctly measured (Eqs.(2,3)). Intrinsically, the nonequilibrium behavior of the system led by the demon is due to the time-irreversibility of the binary relationship between them, which is quantified by the dissipative information (Eq.(1)). Besides, we prove another new fluctuation theorem for this dissipative information (Eq.(6)). This theorem (Eq.(6)) combining with the other new fluctuation theorems (Eq.(4)) for the entropy productions gives a precise quantification of the effect of the demon. An apparent result following these theorems is that there exists an inevitable energy dissipation originated from the positive dissipative information, which leads to the tighter bounds for the work and the heat (Eq.(11)) than that estimated by the ordinary 2nd law. We also suggest a possible realization of the experimental estimation of these work and heat bounds, which can be measured and tested. These results offer a general picture of a large class of the models of the Maxwell's demon.

\emph{Proof of the Fluctuation Theorems--}The probabilities (densities) $p[x(t)|y]$ and $p[x(t)]$ are assumed to be nonnegative and to be normalized, i.e., $p[x(t)|y], p[x(t)]\ge 0$ respectively, $\int p[x(t)|y] D x(t)=1$ and $\int p[x(t)] D x(t)=1$. Besides, we need that the differentials with respect to the time-forward and backward trajectories are equal to each other, i.e., $D x(t)=D \widetilde{x}(t)$. For the entropy productions and the dissipative information in Eqs.(1,2,3), we obtain the equalities,
\begin{equation*}
\begin{matrix}
   &\langle \exp(-\sigma_{X|Y})\rangle=\int dy \int p(y)p[x(t)|y]\frac{p[\widetilde{x}(t)|y]}{p[x(t)|y]}D x(t)=1\\
   &\langle \exp(-\sigma_{X})\rangle=\int p[x(t)]\frac{p[\widetilde{x}(t)]}{p[x(t)]}D x(t)=1\\
   &\langle \exp(-\sigma_{I})\rangle=\int dy \int p[x(t)]p[y|x(t)]\frac{p[y|\widetilde{x}(t)]}{p[y|x(t)]}D x(t)=1\\
\end{matrix}
\end{equation*}
In the last equation for $\sigma_{I}$, by noting the relation in the probabilities that $p[y|x(t)]=\frac{p(y)p[x(t)|y]}{p[x(t)]}$, we have $\sigma_{I}=i-\widetilde{i}=\log \frac{p[y|x(t)]}{p[y|\widetilde{x}(t)]}$. This completes the proof on the new FTs in Eqs.(4,6).

\section{Acknowledgements}
Qian Zeng thanks the support in part by National Natural Science Foundation of China (NSFC-91430217).

\end{document}


\preprint{APS/123-QED}

\title{Supplemental Material}


\begin{abstract}
We give the general discussions and calculations on the experimental estimations of the desired thermodynamical entities of the Szilard's type demon: the heat, the work, the entropy production and the dissipative information. We then show the detailed settings and calculations on the cases of the information ratchet, which can be implemented by a particle system.
\end{abstract}

\maketitle

\section{Experimental Estimations on Thermodynamical Entities}
\subsection{Coarse-Grained Markovian Dynamics of Controlled System}

In real cases, the estimations of the entropy productions (Eqs.(2,3)) and the dissipative information $\sigma_I$ (Eq.(1)) can have difficulties in dealing with the non-Markovian systems, where the results may highly depends on the system trajectories. Here we present a heuristic method for simplifying the estimations on the new bounds of the work and the heat given in Eq.(11). The Szilard's type demon is considered in the following discussion\cite{0,1,2,3,4,5}. For a general description, let us still call the controller ``demon" in both the measurement and feedback control processes, and the controlled system is referred to ``system". We still assume that the controlled system is coupled to a single thermal environmental bath where the temperature is set to $T=1$.

The demon performs control to the system in a finite time $t$. The general settings can be given as follows. At the beginning, the system and the demon are at the states $x_{0}$ and $y$ respectively. The probability of $x_0$ is represented by $p_{X_0}(x_0)$. Then $x_{0}$ and $y$ can be correlated at the initial time with stable conditional probability $p_{X_0|Y}(x_{0}|y)$. The Szilard's type demon can achieve a control $\Gamma(y)$ to change the state from $x_{0}$ to $x_{c}$ immediately right after the initial time. Here, $x_c$ needs not to be equal to $x_0$. Due to this protocol, the state $x_{c}$ is then correlated with $y$ described by the probability $p_{X_c|Y}(x_c|y)$. Besides, the control time can be short enough before the system reacts. In this situation, the demon permutes the probabilities for the states in the distribution $p_{X_0}$ during the control, and turns $p_{X_0|Y=y}$ to be $p_{X_c|Y=y}$ right after the control. That is, the conditional probabilities of the states $p_{X_c|Y}(x_c|y)=p_{X_0|y}(x_0|y)$ if the state is changed from $x_0$ to $x_c$ at a given $y$. Here, more details about the conditional probabilities $p_{X_0|Y}(x_{0}|y)$ and $p_{X_c|Y}(x_{c}|y)$ should be discussed in the measurement and feedback control processes individually.
\begin{enumerate}
  \item In the measurement process where the system is supposed to measure the demon state $y$, the system does not influence the demon state. Thus, the system and the demon are uncorrelated at the initial state before the control (measurement). Based on the settings in the above, we then have
      \begin{equation}
      \label{S.1}
      \begin{cases}\tag{S.1}
      p_{X_0|Y}(x_0|y)=p_{X_0}(x_0), \\
      p_{X_c|Y}(x_c|y)=p_{X_0}(x_0).
      \end{cases}
      \end{equation}
      Besides, the a priori probability of the demon state $p(y)$ should be known for the succeeding calculations.
  \item In the feedback control process, the demon state $y$ is regarded as the observation of the system state $x_0$. For this reason, $y$ is assumed to be initially correlated with $x_0$ with stable probability $p_{Y|X_0}(y|x_0)$ which should be known for the following discussions. According to the Bayes' rule for the probabilities, we have that
      \begin{equation}
      \label{S.2}
      \begin{cases}\tag{S.2}
      p(y)=\sum_{x_0} p_{Y|X_0}(y|x_0)P_{X_0}(x_0),\\
      p_{X_0|Y}(x_0|y)=\frac{1}{p(y)}p_{Y|X_0}(y|x_0)p_{X_0}(x_0), \\
      p_{X_c|Y}(x_c|y)=p_{X_0|Y}(x_0|y).
      \end{cases}
      \end{equation}
      Here, we should emphasize that the conditional probability $p_{X_c|Y}(x_c|y')$ can be unequal to $p_{X_0|Y}(x_0|y')$ at another $y'\neq y$ although $p_{X_c|Y}(x_c|y)=p_{X_0|Y}(x_0|y)$. This is due to concrete control protocols.
\end{enumerate}

After the control, the demon will not disturb the system dynamics in the rest time. The conditional probability of the final system state is represented by $p_{X_t|Y}(x_t|y)$. Driven by the bath, the system dynamics is conditionally Markovian right after the control with the new initial state $x_c$. The Markovian dynamics in continuous time can be given by the following master equation,
\begin{equation}
\label{S.3}
\partial_t p_{X_t|Y}=K p_{X_t|Y},\tag{S.3}
\end{equation}
where $p_{X_t|Y}$ denotes the conditional distribution of the system state $x$ conditioning on the demon state $y$ at time $t$; $K$ denotes the matrix of the transition rates of the dynamics. Then given the initial conditional distribution as $p_{X_c|Y}$, the solution of Eq.(S.3) can be given by
\begin{equation}
\label{S.4}
p_{X_t|Y}=\exp(Kt)p_{X_c|Y}.\tag{S.4}
\end{equation}
The system will go to the equilibrium exponentially fast. If the time $t$ is long enough for the stochastic environmental force to drive the system, then the final conditional distribution of the system state can be approximately taken as the equilibrium distribution $p_{X_t|Y}=p^{ss}_{X|Y}$. Here the the equilibrium distribution satisfies the equation $p^{ss}_{X|Y}=Qp$ with arbitrary initial distribution $p$. Here $Q=\lim_{t\to \infty}\exp(Kt)$ is recognized as the matrix of the transition probabilities in the long time limit, and each columns of $Q$ are exactly equal to the equilibrium distribution $p^{ss}_{X|Y}$. This suggests that for the long time evolution, the coarse-grained dynamics can be approximately given by
\begin{equation}
\label{S.5}
p_{X_t|Y}=Qp_{X_c|Y},\tag{S.5}
\end{equation}
where the transition probability is independent of $x_c$, i.e., $Q(x_t|x_c,y)=p_{X_t|Y}(x_t|y)=p^ss_{X|Y}(x_t|y)$.

The probability of the demon state $y$, denoted by $p(y)$, is assumed to be known in the general discussion. By using $p(y)$, the probabilities of the system states can be evaluated as follows, $p_X(x)=\sum_y p(y)p_{X|Y}(x|y)$. The coarse-grained dynamics of the system is in general non-Markovian. Also, there is a discontinuity in the dynamics at the control time. To calculate the entropy productions and the dissipative information, we can evaluate the probabilities of the coarse-grained trajectory $x(t)=\{x_c, x_t\}$ (the initial state $x_0$ should be known) and the time-reversal trajectory $\widetilde{x}(t)=\{x_t, x_c\}$ right after the control, by using the Markovian nature for the dynamics (Eq.(S.5)). We then have the following probabilities of the trajectories as follows,
\begin{enumerate}
  \item For the measurement process (due to Eq.(S.1)), we have
     \begin{equation}
     \label{S.6}
     \begin{cases}\tag{S.6}
     p[x(t)|y]=p_{X_0}(x_0)p^{ss}_{X|Y}(x_t|y)\\
     p[\widetilde{x}(t)|y]=p^{ss}_{X|Y}(x_t|y)(x_t|y)p^{ss}_{X|Y}(x_c|y),\\
     p[x(t)]=p_{X_0}(x_0)\sum_y p(y)p^{ss}_{X|Y}(x_t|y),\\
     p[\widetilde{x}(t)]=\sum_y p(y)p[\widetilde{x}(t)|y].
     \end{cases}
    \end{equation}
  \item For the feedback control process (according to Eq.(S.2)), we have that
      \begin{equation}
      \label{S.7}
      \begin{cases}\tag{S.7}
      p[x(t)|y]=p_{X_c|Y}(x_c|y)p^{ss}_{X|Y}(x_t|y),\\
      p[\widetilde{x}(t)|y]=p^{ss}_{X|Y}(x_t|y)(x_t|y)p^{ss}_{X|Y}(x_c|y), \\
      p[x(t)]=\sum_y p(y)p[x(t)|y],\\
      p[\widetilde{x}(t)]=\sum_y p(y)p[\widetilde{x}(t)|y].
      \end{cases}
      \end{equation}
  Particularly, for the cases where the final correlation between the system and the demon vanishes, we have that $p^{ss}_{X|Y}(x|y)=p^{ss}_{X}(x)$. Then Eq.(S.7) can be reduced into
      \begin{equation}
      \label{S.8}
      \begin{cases}\tag{S.8}
      p[x(t)|y]=p_{X_c|Y}(x_c|y)p^{ss}_{X}(x_t),\\
      p[\widetilde{x}(t)|y]=p^{ss}_{X}(x_t)p^{ss}_{X}(x_c), \\
      p[x(t)]=p^{ss}_{X}(x_t)\sum_y p(y)p_{X_c|Y}(x_c|y),\\
      p[\widetilde{x}(t)]=p^{ss}_{X}(x_t)p^{ss}_{X}(x_c).
      \end{cases}
      \end{equation}
\end{enumerate}
Then we can insert the probabilities in Eq.(S.6) and Eq.(S.7) (or Eq.(S.8)) into Eqs.(1,2,3) to evaluate the entropy productions and the dissipative information in the measurement and the feedback control processes respectively. Furthermore, Eq.(S.6) and Eq.(S.7) indicate that the thermodynamical entities, which depend on the system trajectories can be easily estimated in the experiments, because the calculation only involves the counting of the probabilities (or relative frequencies) of the states.
\subsection{Heat, Work, Free Energy Change, and Entropy Production}
The initial Hamiltonian of the system can be independent of the current demon state $y$, denoted by $H_0(x)$. The Hamiltonian right after the control can depend on both the states of the system and the demon, and is represented by $H_c(x,y)$. During the control, the Szilard's type demon can change the Hamiltonian fast enough before the system reacts. Then we can assume that during the control the change in the total energy is totally caused by the averaged work performed on the system, that is,
\begin{equation}
\label{S.9}
\langle W_{X|Y}\rangle=\langle H\rangle_c-\langle H\rangle_0,\tag{S.9}
\end{equation}
where $\langle H\rangle_{0}=\sum_{x}p_{X_0}(x)H_0(x)$ and $\langle H\rangle_{c}=\sum_{x,y}p(y)p_{X_c|Y}(x|y)H_c(x,y)$ are the total energies right before and after the control respectively.

The total energy change during the dynamics is recognized as,
\begin{equation}
\label{S.10}
\langle \Delta H\rangle=\langle H\rangle_t-\langle H\rangle_0,\tag{S.10}
\end{equation}
where $\langle H\rangle_{t}=\sum_{x,y}p(y)p_{X_t|Y}(x|y)H_c(x,y)$ is the total energy at the final state.

Then following the thermodynamical 1st law, the averaged heat absorbed by the system can be given by,
\begin{equation}
\label{S.11}
\langle Q_{X|Y}\rangle=\langle \Delta H\rangle-\langle W_{X|Y}\rangle=\langle H\rangle_t-\langle H\rangle_c.\tag{S.11}
\end{equation}

Thus, the true entropy production right after the control can be given by,
\begin{equation}
\label{S.12}
\langle \sigma_{X|Y}\rangle=\langle \Delta s_{X|Y}\rangle-\langle Q_{X|Y}\rangle\ge0.\tag{S.12}
\end{equation}
Here the true entropy change is quantified by
\begin{equation}
\label{S.13}
\langle \Delta s_{X|Y}\rangle=\sum_{x,y}p(y)p_{X_c|Y}(x|y)\log p_{X_c|Y}(x|y)-\sum_{x,y}p(y)p_{X_t|Y}(x|y)\log p_{X_t|Y}(x|y).\tag{S.13}
\end{equation}

By recalling the thermodynamical 1st law, we can rewrite the true entropy production in Eq.(S.12) in terms of the work instead of the heat as follows,
\begin{equation}
\label{S.14}
\langle \sigma_{X|Y}\rangle=\langle W_{X|Y}\rangle- \Delta F\ge0,\tag{S.14}
\end{equation}
In Eq.(S.14), the term $\Delta F$ is known as the effective free energy change during the dynamics, and the 2nd law $\langle \sigma_{X|Y}\rangle\ge 0$ still holds for the work. Then according to Eq.(S.12) and Eq.(14), $\Delta F$ can be given as follows,
\begin{equation}
\label{S.15}
\Delta F=\langle \Delta H\rangle- \langle \Delta s_{X|Y}\rangle,\tag{S.15}
\end{equation}
where the total energy change $\langle \Delta H\rangle$ and the true entropy change $\langle \Delta s_{X|Y}\rangle$ have been given in Eq.(S.10) and Eq.(S.13) respectively.

We should note that in the measurement process the averaged measurement heat can be given as $\langle Q_{mea}\rangle=-\langle Q_{X|Y}\rangle$, and in the feedback control the extracted work can be taken as $\langle W_{ext}\rangle=-\langle W_{X|Y}\rangle$.

\subsection{Dissipative Information}
The dissipative information for the general case has been given by Eq.(1). We take the average of the dissipative information as follows,
\begin{equation}
\label{S.16}
\langle \sigma_{I}\rangle=\sum_{y,x(t)}p(y)p[x(t)|y]\left\{\log\frac{p[x(t)|y]}{p[x(t)]}- \log\frac{p[\widetilde{x}(t)|y]}{p[\widetilde{x}(t)]}\right\}.\tag{S.16}
\end{equation}

Here, for the Szilard's type demon, we can give the explicit expressions of $\langle \sigma_{I}\rangle$ in the measurement and the feedback control processes as follows,
\begin{enumerate}
  \item For the measurement process (Eq.(S.6)), we have
     \begin{equation}
     \label{S.17}
     \langle \sigma_{I}\rangle=I_t-\langle \widetilde{i}\rangle,\tag{S.17}
    \end{equation}
    where
    \begin{equation*}
     \begin{cases}
     I_t=\sum_{y,x_t}p(y)p^{ss}_{X|Y}(x_t|y)\log\frac{p^{ss}_{X|Y}(x_t|y)}{\sum_y p(y)p^{ss}_{X|Y}(x_t|y)},\\
     \langle \widetilde{i}\rangle=\sum_{y,x_c,x_t}p(y)p_{X_c|y}(x_c|y)p^{ss}_{X|Y}(x_t|y) \log\frac{p^{ss}_{X|Y}(x_t|y)(x_t|y)p^{ss}_{X|Y}(x_c|y)}{\sum_yp(y)p^{ss}_{X|Y}(x_t|y)(x_t|y)p^{ss}_{X|Y}(x_c|y)}.
     \end{cases}
    \end{equation*}
    Here $I_t$ is the mutual information which quantifies the final correlation between the system and the demon; and $\langle \widetilde{i}\rangle$ is the averaged dynamical mutual information along the time-reversed trajectories.
  \item For the feedback control process, consider the cases where the final correlation vanishes (Eq.(S.8)), we have that
  \begin{equation}
     \label{S.18}
     \langle \sigma_{I}\rangle=I_c=\sum_{y,x_c}p(y)p_{X_c|Y}(x_c|y)\log\frac{p_{X_c|Y}(x_c|y)}{\sum_y p(y)p_{X_c|Y}(x_c|y)}.\tag{S.18}
    \end{equation}
    Here $I_c$ is the mutual information which quantifies the remained correlation between the system and the demon right after the control.
\end{enumerate}

Although the system trajectories start with the state $x_c$ right after the control but not the initial state $x_0$, the new bounds in Eq.(10) or Eq.(11) still hold for the heat and the work in the case of the Szilard's type demon. This is because the heat $Q_{X|Y}$ is generated along the trajectories $x(t)=\{x_c, x_t\}$, and the dynamics shown in Eq.(S.3) is consistent in time. Consequentially, the dissipative information $\langle \sigma_{I}\rangle$ which is measured along the trajectories $x(t)$ can be used to bound the averaged heat shown in Eq.(10) or Eq.(11). On the other hand, the effective free energy change $\Delta F$ guarantees that the averaged work $\langle W_{X|Y}\rangle$ can be bounded by using the entropy production $\langle \sigma_{X|Y}\rangle$ quantified along the trajectories $x(t)$. Then $\langle W_{X|Y}\rangle$ can be also bounded by using $\langle \sigma_{I}\rangle$ since $\langle W_{X|Y}\rangle- \Delta F=\langle\sigma_{X|Y}\rangle\ge \langle \sigma_{I}\rangle$ (Eq.(8)). This yields the new bounds for the work in Eq.(10) or Eq.(11).

\section{Detailed Settings and Calculations on the Cases of the Information Ratchet}
\subsection{Settings of the Particle System}
A particle is confined in a 1-dimension finite zone ranged from $s=-1$ to $1$. Here $s$ is the position of the particle in the zone. The particle can feel a potential $U(s)$ with two ``flat" wells. The potential has two possible profiles as shown in Fig. 1. The functions of the two profiles can be respectively given by
\begin{equation}
\label{S.19a}
U_0(s)=\begin{cases}\tag{S.19a}
0, &s\in[-1,0),\\
V, &s\in[0,1),
\end{cases}
\end{equation}
and
\begin{equation}
\label{S.19b}
U_1(s)=\begin{cases}\tag{S.19b}
V, &s\in[-1,0),\\
0, &s\in[0,1),
\end{cases}
\end{equation}
where the subscript of the potential profile $0$ or $1$ represents the location of the lower well as on the left or the right half of the zone. Here $V>0$ is referred to potential height.

\begin{figure}[!hbp]
\centering
\includegraphics[height=10cm,width=16cm]{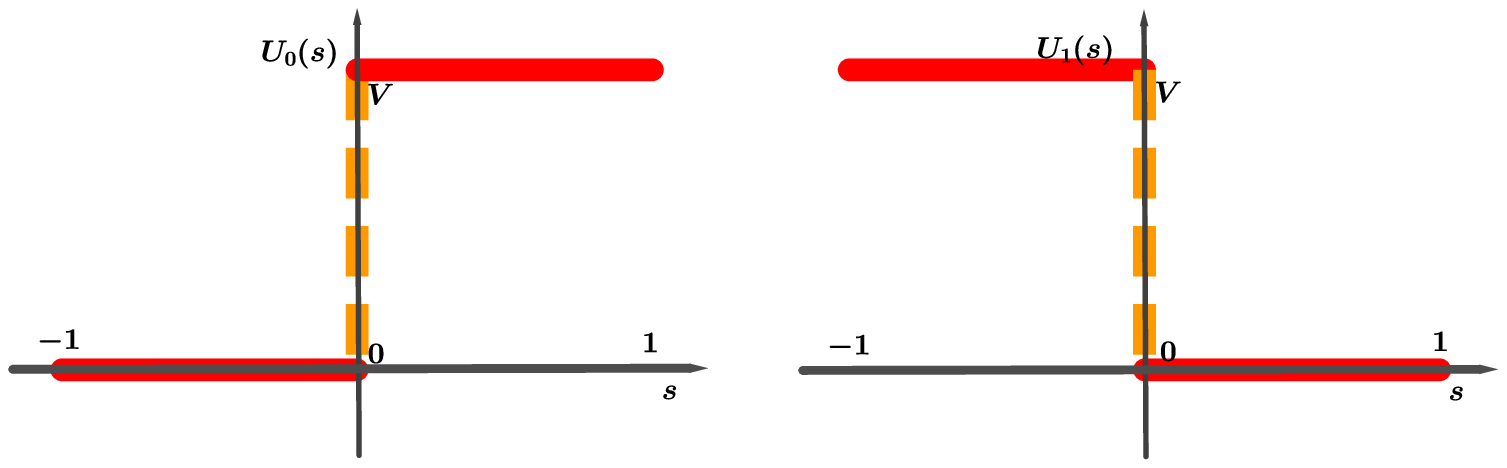}
\caption{The potential profiles given in Eq.(S.19).}
\label{FIG1}
\end{figure}

The particle is driven by a random force $\xi(t)$ resulted from the environmental thermal bath with temperature $T=1$. Consequentially, the particle obeys a Langevin dynamics where the inertia effect can be neglected,
\begin{equation}
\label{S.20}
0=-\gamma\frac{ds}{dt}-\frac{dU(s)}{ds}+\xi(t),\tag{S.20}
\end{equation}
where $-\gamma\frac{ds}{dt}$ is the viscous friction force with the (friction) coefficient $\gamma>0$; $\xi(t)$ is the random force (Gaussian white noise) characterized by the following relations, $\langle\xi(t)\rangle=0$, and $\langle\xi(t)\xi(t')\rangle=2\gamma\beta^{-1}\delta(t-t')$.

If the potential height $V$ satisfies the inequality $V\le1$, then the particle can ``jump" across the barrier and move to either of the wells. Thus, starting from arbitrary position in the zone, the system can reach the equilibrium steady state with the stationary distribution of the position $\pi(s)= Z^{-1}\exp[-U(s)]$, where $Z=\int_{-1}^{1}\exp[-U(s)]ds=1+\exp(-V)$ is the partition function. Then the probabilities that the particle can be found at either of the two wells in the equilibrium steady state can be given by:
\begin{equation}
\label{S.21}
\begin{cases}\tag{S.21}
\text{probability at the lower well, }p_l=\int_{\text{location of lower well}}\pi(s)ds=\frac{1}{1+\exp(-V)},\\
\text{probability at the higher well, }p_h=\int_{\text{location of higher well}}\pi(s)ds=\frac{\exp(-V)}{1+\exp(-V)}.
\end{cases}
\end{equation}

When the particle feels no potential, the relaxation time to reach the equilibrium is approximately the diffusion time $\tau_{0}=2/D$, where $D=\gamma^{-1}$ is the diffusion coefficient. Then the transitions rates of the particle between the two wells can be roughly proportional to the inverse of $\tau_{0}$:
\begin{equation}
\label{S.22}
\begin{cases}\tag{S.22}
\text{rate from the lower to higher well, }r_{l\to h}=v_{0}\exp(-\frac{1}{2} V),\\
\text{rate from the higher to lower well, }r_{h\to l}=v_{0}\exp(\frac{1}{2} V),
\end{cases}
\end{equation}
where $v_{0}=\tau^{-1}_{0}=D/2$. Then the average dwelling time that the particle is at the lower well or the higher well can be given by the inverses of the transition rates as follows,
\begin{equation}
\label{S.23}
\begin{cases}\tag{S.23}
\text{dwelling time at the lower well, }\tau_{l}=\frac{1}{r_{h\to l}},\\
\text{dwelling time at the higher well, }\tau_{h}=\frac{1}{r_{l\to h}},
\end{cases}
\end{equation}

By using the transition rates in Eq.(S.22), we can formulate a series of master equations in the form of Eq.(S.3). These equations can be constructed based on an essential master equation which characterizes the dynamics of the particle as follows,
\begin{equation}
\label{S.24}
\begin{cases}\tag{S.24}
\frac{du_l(t)}{dt}=r_{h\to l}u_h(t)-r_{l\to h}u_l(t),\\
\frac{du_h(t)}{dt}=r_{l\to h}u_l(t)-r_{h\to l}u_h(t),
\end{cases}
\end{equation}
where $u_l(t)$ and $u_h(t)$ ($u_l(t)+u_h(t)=1$) represent the probabilities that the particle is at the lower and the higher wells at time $t$ respectively. In general, given arbitrary initial distribution, the particle finally goes to the equilibrium distribution as $u_l\to p_l$ and $u_h\to p_h$. The corresponding relaxation time can be estimated by,
\begin{equation}
\label{S.25}
\tau_{sys}=\frac{\tau_0}{2\cosh(\frac{1}{2} V)}=\frac{1}{r_{l\to h}+r_{h\to l}}<\tau_{h}<\tau_{l}.\tag{S.25}
\end{equation}
We can see from Eq.(S.25) that the relaxation time $\tau_{sys}$ is less than the average dwelling times at each wells. Thus, it is possible to control the particle with fast operation rate before the particle jumps between wells by switching the potential profiles, and then the particle can goes to the equilibrium steady state in a finite time after the control\cite{6}.

\begin{figure}[!hbp]
\centering
\includegraphics[height=10cm,width=16cm]{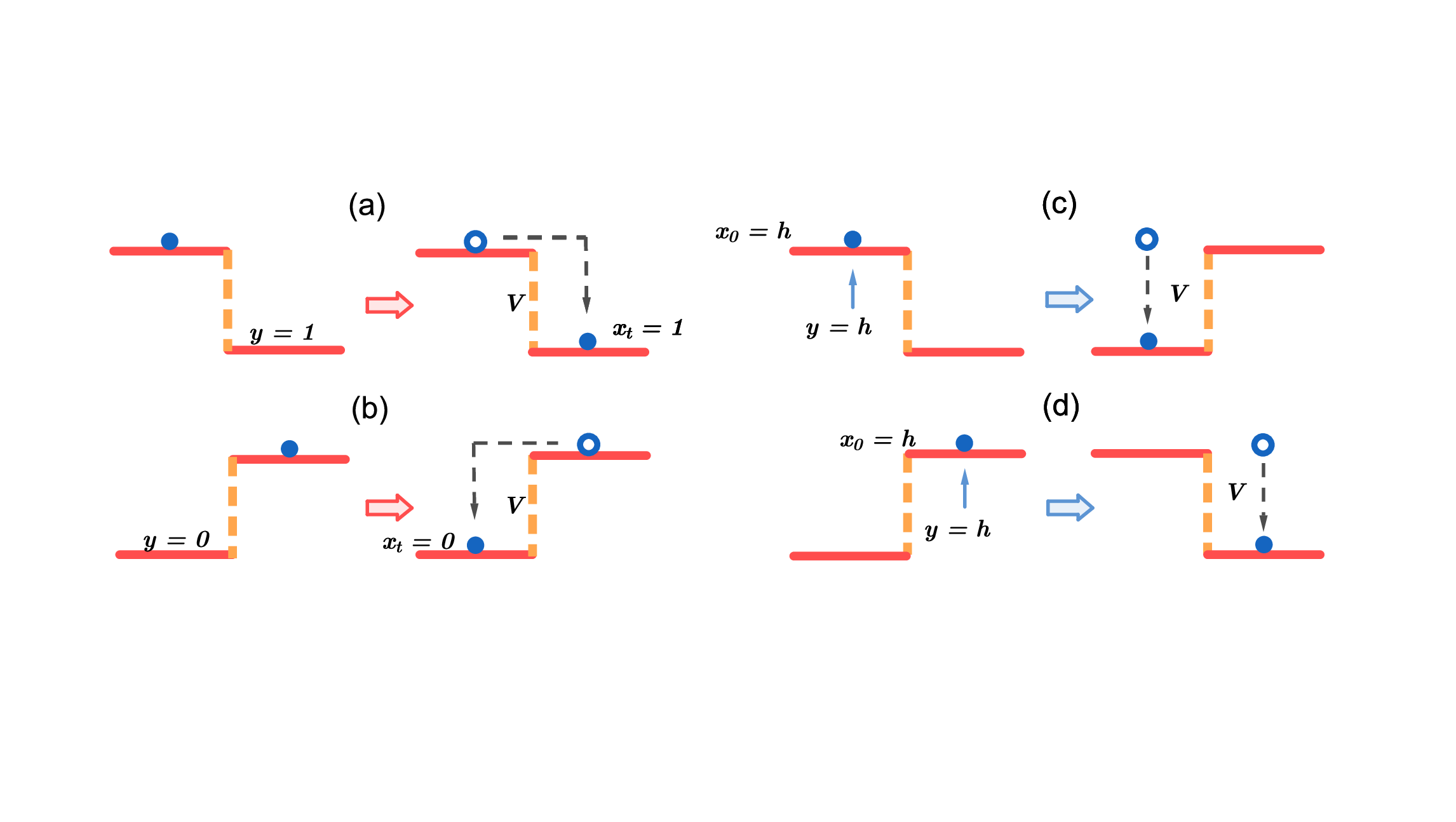}
\caption{The confined particle works as a demon or is controlled by a demon. In (a) and (b), the particle is used as a demon and measures the system state before the control. The system state, denoted by $y$, is represented by the location of the lower well ($0$ or $1$) in the potential. In (a), $y=1$; and In (b), $y=0$. The final state of the particle, denoted by $x_t$, is taken as the observation of $y$. In (c) and (d), the particle is controlled by a demon. The initial state of the particle, denoted by $x_0$, is represented by $h$ or $l$ when the particle is at the higher or the lower well. When spotting the state $x_0=h$, the demon reverses the potential and extracts a positive work of the particle system.}
\label{FIG4}
\end{figure}

\subsection{Information Measurement}
If the particle works as a demon (shown in Fig.2. (a) and (b)), the particle is supposed to measure the state of the controlled system at first. The state of the particle is denoted by $x=0$ or $1$ when at the left or the right well respectively. The system state can be represented by the location of the lower well with the value of $y=0$ or $1$ with equal probability $p(y=0)=p(y=1)=1/2$. The particle is initially under the equilibrium until the system state changes. Correspondingly, the potential profile is reversed by the system immediately and the particle starts measuring the current system state. When the equilibrium is achieved, the final state $x_t$ of the particle is taken as an observation of $y$. The probability of the measurement error can be given by the probability of the particle at the higher well, $p_{X_t|Y}(x_t\neq y|y)=p_h$ (given by Eq.(S.21)). On the other hand, the measurement precision is characterized by the probability of the particle at the lower well, $p_{X_t|Y}(x_t= y|y)=p_l$.

Due to the description of the information measurement, we rewrite the potential profiles in Eq.(S.19) as follows,
\begin{equation}
\label{S.26a}
U(x,0)=\begin{cases}\tag{S.26a}
0, &x=0,\\
V, &x=1,
\end{cases}
\end{equation}
and
\begin{equation}
\label{S.26b}
U(x,1)=\begin{cases}\tag{S.26b}
V, &x=0,\\
0, &x=1,
\end{cases}
\end{equation}

According to the Markovian dynamics given in Eq.(S.24), we can evaluate the probabilities of the particle trajectories by using Eq.(S.6), given the initial potential profile as $U(x,y')$ ($y'=0,1$) in Eq.(S.26). Since the potential profile change does not alter the position of the particle in this case, the particle state $x_c$ right after the control is equal to the initial state $x_0$, i.e., $x_c=x_0$. The probabilities of the trajectories $x(t)=\{x_c,x_t\}$ and the time-reversal trajectories $\widetilde{x}(t)=\{x_t,x_c\}$ are given in Table I and Table II respectively.

On the other hand, the value of $y'$ (in the expression of the initial profile $U(x,y')$) can be taken as the system state in the previous control (measurement) cycle when performing the cyclic control, where the probabilities of $y'$ can be given by $p(y'=0)=p(y'=1)=1/2$. Thus, the final (equilibrium) distribution in the previous cycle, represented by $p_{X_t|Y'}$ is exactly the initial distribution in the current cycle. Then we can rewrite the probability of the initial state $p_{X_0}(x_0)$ into $p_{X_0|Y'}(x_0|y')$. Consequentially, the probability of the state $x_c=x_0$ right after the control can be given by $p_{X_0|Y'}(x_0|y')$. With this thought, the averaged measurement heat (see Eq.(S.11)) in this case can be given by
\begin{equation}
\label{S.27}
\langle Q_{mea}\rangle=\langle H\rangle_c-\langle H\rangle_t=(1/2-p_h)V.\tag{S.27}
\end{equation}
Here, the total energy right after the control is given by $\langle H\rangle_c=\sum_{y',y,x}p(y')p(y)p_{X_0|Y'}(x|y')U(x,y)=V/2$; and the final total energy is given by $\langle H\rangle_t=\sum_{y,x}p(y)p_{X_t|Y}(x|y)U(x,y)=p_hV$.

The entropy change (see Eq.(S.13)) can be given by
\begin{equation}
\label{S.28}
\langle \Delta s_{X|Y}\rangle=S_{X_t|Y}-S_{X_c|Y}=-I_t.\tag{S.28}
\end{equation}
Here the (conditional) entropy right after the control is given by $S_{X_c|Y}=S=-p_l\log p_l-p_h\log p_h$; and the final entropy is given by $S_{X_t|Y}=S-I_t$, where $I_t=\log 2-S\ge0$ is the final mutual information which measures the correlation between the observation $x_t$ and the state $y$.

The dissipative information can be evaluated by inserting the probabilities shown in Table I and Table II into Eq.(S.17). Due to the symmetry of the potential profile in Eq.(S.26), we have the dissipative information as follows,
\begin{equation}
\label{S.29}
\langle\sigma_{I}\rangle=\log\sqrt{2p_l^2+2p_h^2}. \tag{S.29}
\end{equation}


\begin{table}[!ht]
\caption{The probabilities of the particle trajectories in the information measurement, given the initial potential profile as $U(x,0)$}
\begin{ruledtabular}
\begin{tabular}{lllllll}
$x(t)$      & $p[x(t)|y=0]$ & $p[x(t)|y=1]$ & $p[x(t)]$& $p[\widetilde{x}(t)|y=0]$ & $p[\widetilde{x}(t)|y=1]$ &$p[\widetilde{x}(t)]$\\
\hline
$\{0,0\}$      & $p_l^2$ & $p_lp_h$ & $\frac{1}{2}p_l$ & $p_l^2$ &$p_h^2$ & $\frac{1}{2}(p_l^2+p_h^2)$ \\
\hline
$\{0,1\}$      & $p_lp_h$ & $p_l^2$ & $\frac{1}{2}p_l$ & $p_lp_h$ &$p_lp_h$ & $p_lp_h$  \\
\hline
$\{1,0\}$  & $p_lp_h$ & $p_h^2$ & $\frac{1}{2}p_h$ & $p_lp_h$ &$p_lp_h$ & $p_lp_h$  \\
\hline
$\{1,1\}$  & $p_h^2$ & $p_lp_h$ & $\frac{1}{2}p_h$ & $p_h^2$ &$p_l^2$ & $\frac{1}{2}(p_l^2+p_h^2)$
\end{tabular}
\end{ruledtabular}
\end{table}

\begin{table}[!ht]
\caption{The probabilities of the particle trajectories in the information measurement, given the initial potential profile as $U(x,1)$}
\begin{ruledtabular}
\begin{tabular}{lllllll}
$x(t)$      & $p[x(t)|y=0]$ & $p[x(t)|y=1]$ & $p[x(t)]$& $p[\widetilde{x}(t)|y=0]$ & $p[\widetilde{x}(t)|y=1]$ &$p[\widetilde{x}(t)]$\\
\hline
$\{0,0\}$      & $p_lp_h$ & $p_h^2$ & $\frac{1}{2}p_h$ & $p_l^2$ &$p_h^2$ & $\frac{1}{2}(p_l^2+p_h^2)$ \\
\hline
$\{0,1\}$      & $p_h^2$ & $p_lp_h$ & $\frac{1}{2}p_h$ & $p_lp_h$ &$p_lp_h$ & $p_lp_h$  \\
\hline
$\{1,0\}$  & $p_l^2$ & $p_lp_h$ & $\frac{1}{2}p_l$ & $p_lp_h$ &$p_lp_h$ & $p_lp_h$  \\
\hline
$\{1,1\}$  & $p_lp_h$ & $p_l^2$ & $\frac{1}{2}p_l$ & $p_h^2$ &$p_l^2$ & $\frac{1}{2}(p_l^2+p_h^2)$
\end{tabular}
\end{ruledtabular}
\end{table}

\subsection{Work Extraction}
We use a demon to extract positive work from the particle system (shown in Fig.2. (c) and (d)). In this case, the state of the particle can be denoted by $x=l$ or $h$ when at the lower or the higher well respectively. Initially, the particle is under equilibrium. he demon measures the state of the particle at first and obtains the observation $y$. The demon plays the feedback control according to the observation $y$. When the particle is observed to be at the higher well, the demon reverses the potential immediately and extracts an amount of work $W_{ext}=V$. After the control, the demon does nothing until the particle goes to the equilibrium again. For a practical thought, the demon's measurement can have a random error and this error certainly lowers the efficiency of the work extraction. Here we simply assume that the measurement error occurs with stable probability $p_{Y|X_0}(y\neq x_0|x_0)=\epsilon$.

Due to the description of the work extraction, we rewrite the potential profiles in Eq.(S.19) as follows,
\begin{equation}
\label{S.30}
U(x)=\begin{cases}\tag{S.30}
0, &x=l,\\
V, &x=h.
\end{cases}
\end{equation}
We can see from Eq.(S.30) that the potential profile in the work extraction depends no more on the demon state $y$. Thus, the demon does not change the Hamiltonian in this case. The particle state $x_c$ right after the control can be written as the function of the initial state $x_0$ and the demon's observation $y$ as follows,
\begin{equation}
\label{S.31}
x_c=\begin{cases}\tag{S.31}
l, &x_0=l,y=l,\\
l, &x_0=h,y=h,\\
h, &x_0=l,y=h,\\
h, &x_0=h,y=l.
\end{cases}
\end{equation}
We can see from Eq.(31) that the state $x_c=h$ only if a measurement error $y\neq x_0$ occurs. Thus, the probability of $x_c=h$ is equal to the error probability, $p_{X_c}(x_c=h)=\epsilon$; and the probability of $x_c=l$ is equal to the precision probability, $p_{X_c}(x_c=l)=1-\epsilon$. The final (equilibrium) distribution of the state only depends on the potential profiles in Eq.(S.30) and is independent of the demon' observation $y$.

According to Eq.(S.31) and the relationships between the probabilities given in Eq.(S.2), we can evaluate the averaged work by using Eq.(S.9) as follows,
\begin{equation}
\label{S.32}
\langle W_{ext}\rangle=\langle H\rangle_0-\langle H\rangle_c=(p_h-\epsilon)V.\tag{S.32}
\end{equation}
Here, the initial total energy is given by $\langle H\rangle_0=\sum_xp_{X_0}(x)U(x)=p_hV$; and the total energy right after the control is given by $\langle H\rangle_c=\sum_{x,y}p(y)p_{X_c}(x|y)U(x)=\sum_{x}p_{X_c}(x)U(x)=\epsilon V$.

The effective free energy difference can be given by Eq.(S.15)
\begin{equation}
\label{S.33}
\Delta F=\langle \Delta H\rangle- \langle \Delta s_{X|Y}\rangle=-I_0.\tag{S.33}
\end{equation}
Here the total energy change $\langle \Delta H\rangle=0$ because the demon does not change the Hamiltonian; the entropy change $\langle \Delta s_{X|Y}\rangle=-I_0$. Here the mutual information $I_0=S_Y-S_\epsilon\ge0$ represents the initial correlation between the demon and the particle, where the Shannon entropies $S_Y=-p_y\log p_y-(1-p_y)\log(1- p_y)$ and $S_\epsilon=-\epsilon\log \epsilon-(1-\epsilon)\log (1-\epsilon)$, with $p_y=p_l(1-\epsilon)+p_h\epsilon$ representing the probability of the observation $y=l$.

The dissipative information in this case can be given by Eq.(S.18) as $\langle \sigma_{I}\rangle=I_c$, where the mutual information $I_c$ measures the remained correlation between the system and the demon right after the control.